\documentclass[useAMS]{mn2e}
\usepackage{graphicx}  % for includegraphicsx
\usepackage{mnras_cite}

\usepackage{dcolumn}  % Align table columns on decimal point
\usepackage{bm}       % bold math
\usepackage{amssymb,amsmath}
\usepackage{color}

\newcommand{\bea}{\begin{eqnarray}}
\newcommand{\eea}{\end{eqnarray}}
\newcommand{\zbar}{\bar{z}}

\newcommand{\beq}{\begin{equation}}
\newcommand{\eeq}{\end{equation}}

\newcommand{\beqa}{\begin{eqnarray}}
\newcommand{\eeqa}{\end{eqnarray}}
\newcommand{\camb}{{\tt CAMB$\_$sources}}

\def\Mpc{\, h^{-1} \, {\rm Mpc}}

\def\kvecMpc{\, h \, {\rm Mpc}^{-1}}

\def\la{\mathrel{\mathpalette\fun <}}
 \def\ga{\mathrel{\mathpalette\fun >}}
\def\fun#1#2{\lower3.6pt\vbox{\baselineskip0pt\lineskip.9pt
\ialign{$\mathsurround=0pt#1\hfil##\hfil$\crcr#2\crcr\sim\crcr}}}

\begin{document}

\title[]
{Recovering 3D clustering information with angular correlations}
\author[Asorey et al.]
{\parbox{\textwidth}{Jacobo Asorey$^1$, Martin Crocce$^1$, Enrique
    Gazta\~naga$^1$, Antony Lewis$^2$
%\footstar{www.cosmologist.info}
}	\vspace{0.4cm}\\
$^1$Institut de Ci\`encies de l'Espai (IEEC-CSIC),  E-08193 Bellaterra (Barcelona), Spain \\
$^2$Department of Physics \& Astronomy, University of Sussex, Brighton BN1 9QH, U.K. \\
}

\date{\today}
\pagerange{1--10} \pubyear{2012}
\maketitle

\begin{abstract}

We study how to recover the full 3D clustering information of $P(\vec{k},z)$, including redshift
space distortions (RSD),  from 2D tomography using the angular auto and cross spectra of
different redshift bins $C_\ell(z,z')$.  We focus on quasilinear scales where the
minimum scale $\lambda_{min}$ or corresponding maximum wavenumber $k_{max}= 2\pi/\lambda_{min}$
is targeted to be between $k_{max}=\{0.05-0.2\} \kvecMpc$.
For spectroscopic surveys,
we find that we can recover the full 3D clustering information
when the redshift bin width $\Delta z$ used in the 2D
tomography is similar to the targeted minimum scale, i.e.
$\Delta z \simeq \{0.6-0.8\} \,\lambda_{min} H(z)/c$ which corresponds
to $\Delta z \simeq 0.01-0.05$ for $z<1$.
This value of  $\Delta z$ is optimal in the
sense that larger values of $\Delta z $ lose information, while
smaller values violate our minimum scale requirement.
For a narrow-band photometric survey, with photo-z error $\sigma_z=0.004$,
we find almost identical results to the spectroscopic survey
because the photo-z error is smaller than the optimal bin width
$\sigma_z<\Delta z$. For a typical broad-band photometric  survey with
$\sigma_z=0.1$, we have that $\sigma_z>\Delta z$ and most
radial information is intrinsically lost.  The remaining  information
 can  be recovered from the 2D tomography if we use $\Delta z \simeq
 2\sigma_z$. While 3D and 2D analysis are shown here to be equivalent,
the advantage of using  angular positions and redshifts
is that we do not need a fiducial cosmology to convert to 3D
coordinates. This avoids assumptions and marginalization over the
fiducial model. In addition, it becomes straight forward to combine RSD, clustering
and weak lensing in 2D space.
\end{abstract}


\section{Introduction}\label{sec:introduction}

In recent years, galaxy redshift surveys have provided new
 information about the cosmological model of our Universe, in pace with precision cosmology from other probes like CMB and type Ia Supernovae. We are now entering
exciting times for cosmology, when surveys will go deeper and wider with increasing
number of galaxy positions in each catalogue. With deep surveys we can use weak lensing information 
to improve constraints on cosmological parameters and also trace directly the dark matter distribution at large scales. Theoretical analysis of weak lensing (WL) is usually made through a 2D (angular) analysis of the measured galaxy shear maps. Future surveys will have less
shot noise, allowing for more freedom in how we break the sample
into multiple redshift shells, so that galaxy correlations can also be measured in and between shells.
In doing angular correlations, we are projecting all the radial information within
each redshift bin. But if we are able to use very thin radial shells,
we can maybe recover the radial information using the angular
cross-correlations between all the redshift bins (see \pcite{1206.3545} for a
related idea). This is what we want to investigate in this paper.

This goal is also connected to recent studies of galaxy surveys
using a combination of redshift space distortions (RSD) and WL
galaxy-shear and shear-shear correlations. These allow measurements of
galaxy bias and the breaking of degeneracies between growth history and cosmic history, as has been
recently proposed \cite{1109.4852,1112.4478}.RSD are usually studied in 3D,
which complicates a joint analysis with
WL which is usually 2D (see \pcite{kitching2011} for a comparative
analysis with 3D cosmic shear).   If we could study RSD in 2D without loss of
information,  then it would be possible to do a joint analysis of both
probes using only angular correlations with the corresponding simplification
in the covariance analysis.

Observations directly probe redshifts and angular coordinates on the sky.
Doing an angular analysis therefore does not require any prior knowledge
of the cosmological model, while for doing 3D analysis we have to
assume a fiducial cosmology to convert to comoving spatial coordinates.
This then requires modelling the Alcock-Paczynski effect when fitting different models to our
observables \cite{alcockpaczynski}. As the transformation is redshift dependent one has
to make sure that this procedure is not biasing the parameter constraints. 
If the theoretical prediction for the correlations in angle and redshift can be calculated for each model, 
an angular analysis relating directly to the observables is much more direct.

The final goal of this paper is to analyze the bin optimization that
allows us to recover the 3D constraints on clustering using a 2D
tomographic approach. We have studied this in the framework of several idealized surveys: a
spectroscopic survey in a  redshift range similar to SDSS redshift
range; a survey with photometric redshifts from a camera with narrow-band filters like the camera that Physics of
the accelerating Universe Survey (PAU)\footnote{www.pausurvey}; and finally, a survey with redshifts obtained from photometry
with broadband filters, in a redshift range similar to Dark Energy
Survey (DES)\footnote{www.darkenergysurvey.org}. For the
three surveys we have analyzed a bias fixed model, constraining
$\Omega_m$. In addition, in the spectroscopic survey we have also
studied the standard RSD constraints on the bias $b$ and growth index $\gamma$.	

In section \ref{sec:methodology} we describe galaxy surveys, parameters considered in the analysis and a description of the observables. In section {\ref{sec:results}} we show the constraints obtained in our analysis for the different surveys described above. Finally, we summarize all the results in section \ref{sec:conclusions} with the conclusions.

\section{Methodology}\label{sec:methodology}

The goal of this paper is to show under which conditions, if any, one can recover the full 3D clustering information from a tomography study. By this we mean a combination of all the auto and cross angular spectra after the survey volume has been divided in a set of consecutive redshift bins.
The angular spectra within each bin will include information mainly from transverse modes, while cross correlations between different shells accounts for radial modes with scales comparable to the bin separation.

We investigate this idea in the context of a spectroscopic survey as well as two photometric surveys with different accuracies in the redshift determination. In what follows we describe these ``typical'' surveys, the assumed galaxy samples, the observables considered and the figures of merit used to compare 3D and 2D tomography results.

Throughout the paper we used
\camb\footnote{camb.info/sources} \cite{2011arXiv1105.5292C} to
compute the matter 3D power spectra as well as the angular power
spectra, including cross correlations between radial bins.

\subsection{Fiducial surveys and galaxy samples}
In this section we describe our fiducial surveys and galaxy samples. We characterize them by a redshift range, a given accuracy of redshift measurements, a galaxy redshift distribution and bias.

In all cases we assume a full sky coverage. In ideal conditions this implies that the covariance matrix of observables such as $C_\ell$ is diagonal in $\ell$ (but notice that this assumption is not expected to change the conclusions of this paper). In all three surveys the overall redshift distribution of galaxies per ${\rm deg}^2$ is taken as,
\begin{equation}
\frac{dN}{dzd\Omega}=N_{gal}  \left(\frac{z}{0.55}\right)^{2}e^{-\left(\frac{z}{0.55}\right)^{1.5}}
\label{eq:galaxyselectionfunction}
\end{equation}
which is typical of a flux-limited sample with a magnitude cut
at $i_{AB}<24$. In Eq.~(\ref{eq:galaxyselectionfunction}) $N_{gal}$ is a normalization related to the total number of galaxies per square degree under consideration.

\begin{table}
{\center
\begin{tabular}{cc}
Case &$n(r)$ ($h^{3} \, {\rm Mpc}^{-3}$) \\ \\
\hline
Low  Shot-Noise &$3.14\cdot 10^{-3}$ \\
High Shot-Noise &$6.89\cdot 10^{-4}$\\
\hline \\
\end{tabular}
\caption{Comoving galaxy number densities at $z=0.55$ assumed in this paper for the spectroscopic and narrow-band photometric surveys. Case 1 corresponds to a low shot noise level ($n P_{gal} \sim 2\%$, where $P_{gal}$ is the monopole of the galaxy spectrum at $z=0.55$ and $k=0.1 \kvecMpc$) while Case 2 corresponds to a high shot noise level ($n P_{gal} = 10\%$) .}
\label{table:Numgal}
}
\end{table}

\begin{figure}
\begin{center}
\includegraphics[width=0.4\textwidth]{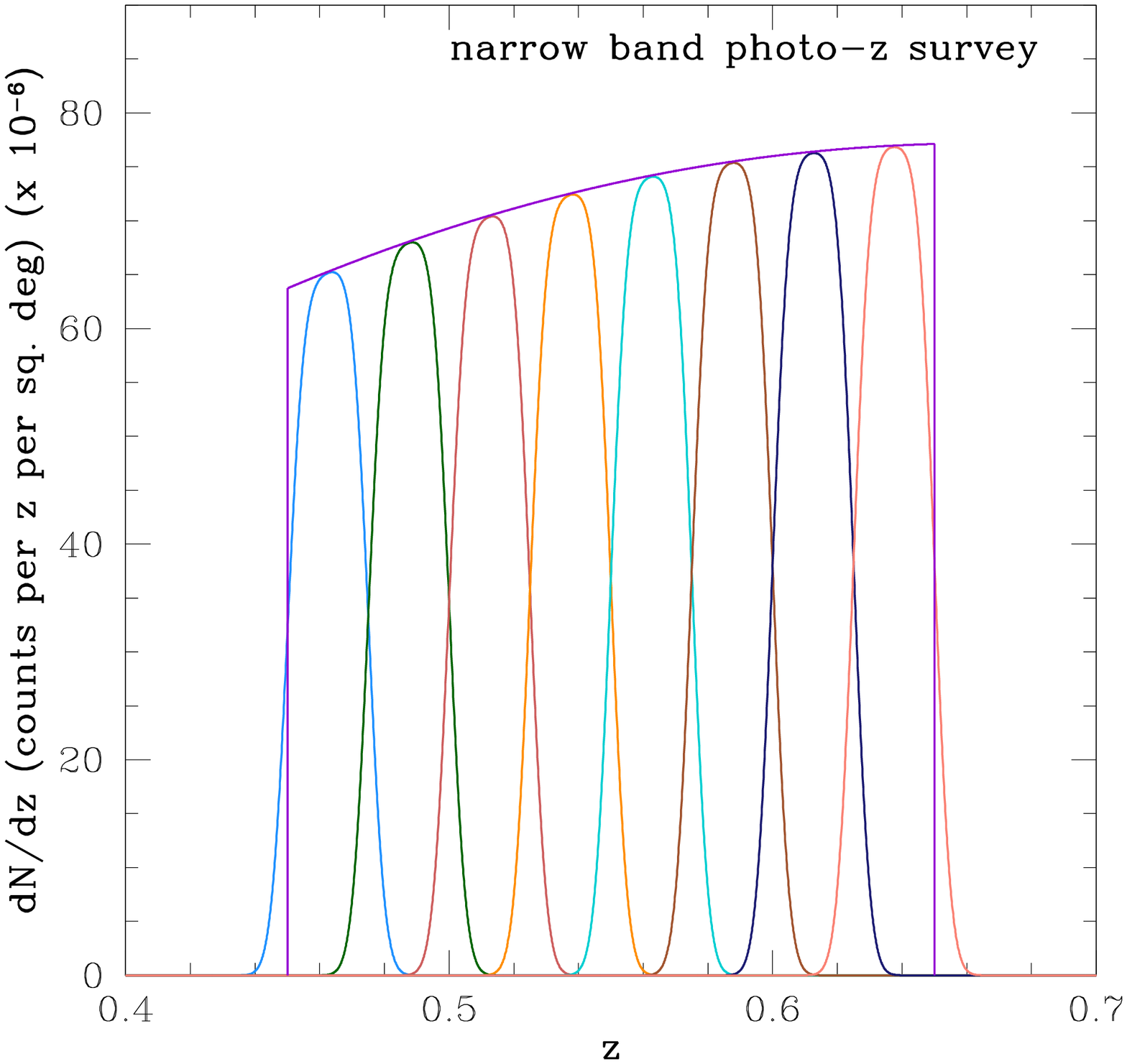}
\includegraphics[width=0.4\textwidth]{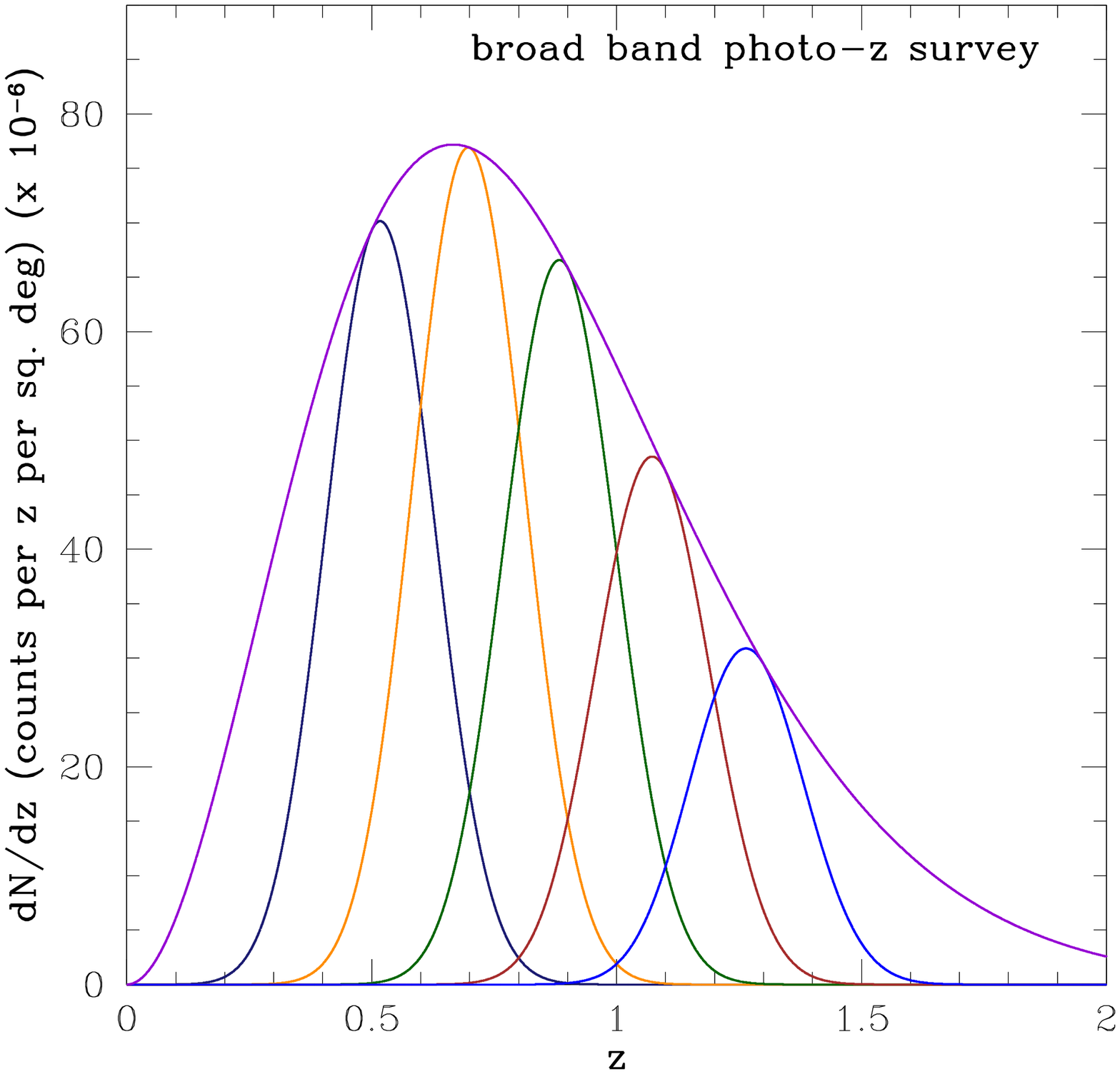}
\caption{Top panel shows the redshift distribution in the
  spectroscopic and narrow band photo-z survey (violet). For the
  narrow band case we show how the true redshift distributions given by Eq.~(\ref{eq:galaxyselectionfunction}) look like if we divide the volume in eight consecutive redshift bins. Bottom panel shows the same but for a broadband photometric survey divided in five bins.}
  \label{fig:dNdz}
\end{center}
\end{figure}

\subsubsection{Spectroscopic survey}
\label{subsec:zsurveys}

Our benchmark spectroscopic survey has radial positions
given by true redshifts (i.e. $\sigma_z=0$ in the formulation below) and a redshift range $0.45<z<0.65$.
Hence for the 2D tomography of this survey we use top hat
bins\footnote{To satisfy differentiability requirements at the edges
  we use in practice $\phi(z)\propto \exp{\left[-((z-\bar{z})/(\Delta
      z /2))^{20}\right]}$, where ${\bar z}$ is the mean redshift of
  the bin and $\Delta z$ the full width.} to compute angular power spectra.
In Table \ref{table:BinConfiguration} we show the different bin configurations considered, characterized by the number of bins in which we divide the survey volume and their width.
Provided with the narrow redshift range we can assume that the bias does not evolve, hence we take $b=2$ throughout.

Lastly we discuss two cases for this survey, one where shot-noise is
non-negligible and another where it is a sub-dominant source of
error. These cases are detailed in Table\ref{table:Numgal}, and for
the redshift range under consideration imply $9$M and $40$M galaxies
respectively (assuming full sky surveys).

\subsubsection{Narrow Band Photometric survey}
\label{subsec:NBphotsurveys}

This case intends to be representative of a configuration such as the one proposed for the PAU survey
where a set of narrow band filters is
expected to deliver ``low-resolution'' spectra in a redshift range
actually broader than the one considered here \cite{Benitez2009,1109.4852}.
Hence our narrow-band photo-z survey has accurate photometric redshifts of $\sigma_z=0.004$, in the same redshift range of the spectroscopic case ($0.45 < z < 0.65$). The bias ($b=2$) and the shot-noise cases considered match those of Sec.~\ref{subsec:zsurveys}  (and are given in Table~\ref{table:Numgal}).

In turn the bin configurations assumed for the 2D tomography  are also the same as for the spectroscopic survey given in Table \ref{table:BinConfiguration}, but with bin limits that now refer to photometric redshifts. Thus the true redshift distribution of galaxies in each bin is no longer a  top hat, but rather has a small overlap with the nearest neighbouring bins due to the photo-z error, as described in Eq.~(\ref{eq:nz}) below. In the top panel of Fig.~\ref{fig:dNdz} we show this effect for the particular case of $8$ bins.

\subsubsection{Broad Band Photometric survey}
\label{subsec:BBphotsurveys}

On the other hand we consider a photometric survey that uses broad-band filters such as DES \footnote{www.darkenergysurvey.org}, Pan-Starrs \footnote{pan-starrs.ifa.hawaii.edu} or the future imaging component of Euclid \footnote{www.euclid-imaging.net}. These surveys are expected to achieve photometric redshift estimates with accuracies $\sigma_z \sim 5\% / (1+z)$ \cite{Banerji2008,2011MNRAS.415.2193R}.
In what follows we do not consider a possible redshift evolution of the photometric error but instead assume a conservative value of $\sigma_z = 0.1$.

Typically optical photo-z surveys are fainter and sample a much larger number of galaxies than spectroscopic ones, hence we assume a broader redshift range, $0.4<z<1.4$, and
only a low shot-noise case as given in Table~\ref{table:Numgal}. For the redshift range assumed this implies $\sim 150\times 10^6$ galaxies. Table \ref{table:BinConfigurationDES} show the bin configurations we have considered for this case.
While in the previous cases we have assumed the bias is constant with redshift (because of the narrow redshift range), for the broadband photometric survey we introduce an evolution following \cite{Fry1996},
\beq
b(z)=1+(b_\star-1)\frac{D(z_{\star})}{D(z)}
\eeq
where $b_{\star} = 2$ is the bias at $z_\star = 1 $. In turn for the evolution of bias we have always assume the fiducial cosmology.

\begin{table}
{\center
\begin{tabular}{ccc}
Number of bins & $\Delta z$ &$\Delta r$ ($h^{-1} \, {\rm Mpc}$)\\ \\
\hline
1&  0.20   & 468      \\
4&  0.05   & 113 - 122\\
8&  0.025  & 56 - 61  \\
16& 0.0125 & 28 - 31  \\
20& 0.010  & 22 - 25  \\
\hline \\
\end{tabular}
\caption{Bin configurations used for the 2D tomography in the case of the spectroscopic and the narrow band photometric survey in a redshift range of $0.45<z<0.65$. We show the number of radial bins and their range of widths in redshift and comoving distance.}
\label{table:BinConfiguration}
}
\end{table}

\begin{table}
{\center
\begin{tabular}{ccc}
Number of bins & $\Delta z$ &$\Delta r$ ($h^{-1} \, {\rm Mpc}$) \\ \\
\hline
4&0.25 & 398 - 592\\
5&0.20  & 315 - 480 \\
6&0.167 & 260 - 404\\
7&0.143  &221 - 348 \\
8&0.125 &193 - 306 \\
9&0.111  &171 - 273 \\
10&0.10 &153 - 246\\
\hline  \\
\end{tabular}
\caption{Bin configurations considered for a broadband photometric survey within a redshift range $0.4<z<1.4$. We show the number of radial bins and their range of widths in redshift and comoving distance.}
\label{table:BinConfigurationDES}
}
\end{table}

\subsection{Spatial (3D) power spectrum}

Since we are only interested in quasi-linear scales we assume the following simple model for the 3D galaxy power spectrum in redshift space,
\beq
P_g(k,\mu,z)=(b+f\mu^2)^2\, D^2(z) \,P_0(k)e^{-k^2{\sigma_t^2}(z)\mu^2},
\label{eq:pk3D}
\eeq
where $P_0$ is the linear spectrum at $z=0$ (properly normalized), $D(z)$ is the linear growth factor and the remaining
amplitude depends on the bias $b(z)$ and the linear growth rate $f(z) \equiv d\ln{D}/ d\ln a$.
The Gaussian cut-off accounts for the fact that the radial information might be diluted due to photometric redshift errors $\sigma_z$ \footnote{This expression is correct as long as the distribution of photometric errors is Gaussian, as we assume throughout this paper.}. In Eq.~(\ref{eq:pk3D}) this redshift error propagates to scales through
$\sigma_t(z)=c \,\sigma_z/H(z)$. Notice that $\sigma_t$ depends also on the cosmic history. This should be taken into account when constraining relevant cosmological parameters (e.g. $\Omega_m$).		

For a spatial analysis the measured 3D power spectrum depends on the
cosmological model assumed to convert redshift and angles to
distances. Hence for every model to be tested against the data one
must perform a new measurement. This process is very costly. Instead one
can choose a {\it reference cosmological model} where the measurement
is done once, and then transform the model prediction to this
reference frame \cite{alcockpaczynski}.

Let us call $P^{obs}(k,\mu)$ the power spectrum measured in the {\it reference} cosmology and $P^{mod}(\tilde{k},\tilde{\mu})$ the model prediction at the point in cosmological parameter space being tested. The transformation of distances and angles from the cosmological model being tested $(\tilde{k}, \tilde{\mu})$ to those in the reference model $(k,\mu)$ is done through the scaling factors
\beq
c_{\parallel}=\frac{H(z)}{H^{mod}(z)} \ \ {\rm ;} \ \
c_{\perp}=\frac{d_A^{mod}(z)}{d_A(z)} ,
\eeq
as $\tilde{k}_\parallel = k_\parallel / c_\parallel$ and $\tilde{k}_\perp = k_\perp / c_\perp$, where $\parallel$ indicates modes parallel to line of sight and $\perp$ perpendicular. The Hubble parameter and the angular diameter distances are given by
\bea
\indent H(z) \! & \! = \! & \! 100h\sqrt{\Omega_m(1+z)^3+\Omega_{DE}(1+z)^{-3(1+w)}}\\
d_A(z) \! & \! = \! & \! \frac{\int_0^{z}{\frac{c\ dz'}{H(z')}}}{1+z}.
\eea
From the above one trivially finds,
\bea
\indent \tilde{k} &=& k \, \sqrt{(1-\mu^2) c_{\perp}^{-2}+\mu^2 c_{\parallel}^{-2}} \\
\tilde{\mu} &=& \mu \, c_\parallel^{-1} /\sqrt{(1-\mu^2) c_{\perp}^{-2}+\mu^2 c_{\parallel}^{-2}}.
\eea
In addition the power spectrum is sensitive to the volume element. Thus we
must re-scale $P^{mod}$ by the differential volume element with respect to the reference cosmology : $c^2_{\perp} c_{\parallel}$.
Lastly, following \cite{tegmark1997} and \cite{seo2003} we construct the $\chi^2$ for each radial bin $i$
as,
\bea
\chi^2_{3D}(i) \!\!\! & \!\!\! = \!\!\! & \!\!\!\int_{k_{min}}^{k_{max}} \frac{dkk^2}{8\pi^2}\int_{-1}^{1}{d\mu }\;{\rm Cov}^{-1}_{\rm eff}(k,\mu)\left(P_g^{obs}(k,\mu,z_i)\right. \nonumber\\
&-&\!\!\left. \frac{1}{c_{\parallel}c_{\perp}^2}P_g^{mod}(\tilde{k},\tilde{\mu},z_i)\right)^2
\label{eq:3Dchi2}
\eea
where ${\rm Cov}^{-1}_{\rm eff}$ is defined for every bin $i$ according to,
\beq
{\rm Cov}^{-1}_{\rm eff} (k,\mu)=\int_{r_{min}(i)}^{r_{max}(i)}\!\!\!d^3r \left(\frac{\bar{n}(r)}{1+\bar{n}(r)P_g^{obs}(k,\mu,\zbar_i)}\right)^2.
\label{eq:effVolume}
\eeq
This is where the covariance of the power spectra is accounted for, which we assume to be diagonal in $k$. It has contributions from both sample variance and shot noise.
In Eqs.~(\ref{eq:3Dchi2},\ref{eq:effVolume}) $P^{obs}$ is the measured spectra in the chosen reference cosmology which we take as our fiducial cosmological model introduced in Sec.~(\ref{sec:cosmologicalmodel}).

For the spectroscopic survey we assume that bins are uncorrelated. Thus the total $\chi^2$ is given by,
\beq
\chi^2_{3D}=\sum_i{\chi^2_{3D}(i)},
\label{eq:3Dtotalchi2}
\eeq
where the sum runs over all the bins considered.

\subsection{Angular (2D) power spectrum}
\label{sec:2Dpower}

In our 2D analysis we consider the exact computation of the angular power spectrum of projected overdensities
in a radial shell,
%\cite{fisher94},
\beq
C_\ell^{ii} = \frac{2}{\pi} \int dk\; k^2 P_0(k) \left(\Psi_l^i(k)+\Psi^{i,r}_l(k)\right)^2
\label{eq:cl}
\eeq
where
\beq
\Psi_\ell^i(k) = \int dz \; \phi_i(z) b(z) D(z) j_\ell(k r(z))
\eeq
is the kernel function in real space and
%$k$ if we do not consider redshift space distortions and
\bea
\Psi^{i,r}_\ell(k)&=&  \int dz \; \phi_i(z) f(z) D(z) \left[ \frac{2l^2+2l-1}{(2\ell+3)(2\ell-1)} j_\ell(kr)
\right. \nonumber \\
&-& \left.\frac{\ell(\ell-1)}{(2\ell-1)(2\ell+1)} j_{\ell-2}(kr)\right.\nonumber\\
&-&\left. \frac{(\ell+1)(\ell+2)}{2\ell+1)(2\ell+3)} j_{\ell+2}(kr) \right].
\eea
should be added to $\Psi^{i}_{\ell}$ if we also include the linear
Kaiser effect \cite{FisherScharfLahav1994,nikhil}. In turn, photo-z effects are included through the radial selection function $\phi(z)$, see below. This model then has the same assumptions as the 3D spectrum from Eq.~(\ref{eq:pk3D}).

Notice that in Eq.~(\ref{eq:cl}) we are only considering density and redshift space
distortions terms. We are neglecting
General Relativity (GR) effects as well as velocity and lensing terms,
which are in our cases subdominant to the ones considered. Nonetheless the
framework of angular auto and cross correlations could easily
include these effects when required \cite{PhysRevD84063505,2011arXiv1105.5292C}.

There are $N_z$ angular power spectrum, one per radial bin. But if we want to study all the clustering information we should add to  our observables the $N_z(N_z-1)/2$ cross-correlations between different redshift bins. These are given by
\beq
C_\ell^{ij} = \frac{2}{\pi} \int dk\; k^2 P(k) \left(\Psi_\ell^i(k)+\Psi^{i,r}_\ell(k)\right) \left(\Psi_\ell^j(k)+\Psi^{j,r}_\ell(k)\right)
\label{eq:clcross}
\eeq
Therefore, we are considering $N_z(N_z+1)/2$ observable angular power spectra when reconstructing clustering information from tomography using $N_z$ bins.

\subsubsection{Radial selection functions}

The radial selection functions $\phi_i$ in Eqs.~(\ref{eq:cl},\ref{eq:clcross}) are the probability to include a galaxy in the given redshift bin. Therefore, they are the product of the galaxy redshift distribution and a window function that depends on selection characteristics (e.g binning strategy),
\beq
\phi_i(z)=\frac{dN_{g}}{dz}W(z)
\eeq
where $dN_{g}/dz$ is given by Eq.~(\ref{eq:galaxyselectionfunction}). We consider two different $W(z)$ depending on the kind of redshifts estimation. In an spectroscopic redshift survey $W(z)$ is a top hat function with the dimensions of the redshift bin. On the other hand, if we include the effect of photo-z then
\beq
W_i(z)=\int{dz_pP(z|z_p)W_i(z_p)},
\eeq
where $z_p$ is the photometric redshift and $P(z|z_p)$ is the probability of the true redshift to be $z$ if the photometric estimate is $z_p$. For the photometric surveys we assume a top-hat selection $W(z_p)$ in photometric redshift and that $P(z|z_p)$ is gaussian with standard deviation $\sigma_z$. This leads to,
\beq
\phi_i(z)\propto\frac{dN_{g}}{dz}\left( {\rm erf}\left[\frac{z_{p,max}-z}{\sqrt{2}\sigma_z}\right]-{\rm erf}\left[\frac{z_{p,min}-z}{\sqrt{2}\sigma_z}\right]\right)
\label{eq:nz}
\eeq
where $z_{p, min}$ and $z_{p, max}$ are the (photometric) limits of each redshift bin considered.
In the equation above and throughout this paper we assume $\sigma_z$ is constant in redshift.

\subsubsection{Covariance matrix of angular power spectra}
The covariance between angular spectra of redshift bins $ij$ and redshift bins $pq$ is given by
\beq
{\rm Cov}_{\ell,(ij)(pq)}=\frac{C_{\ell}^{obs, ip} C_{\ell}^{obs, jq}+
  C_{\ell}^{obs, iq} C_{\ell}^{obs, jp}}{N(l)}
\label{eq:CovObs2D}
\eeq
where $N(\ell)=(2\ell+1)\Delta\ell f_{sky}$ is the number of transverse modes at a given $\ell$ and $\Delta \ell$ is typically chosen to make ${\rm Cov}$ block-diagonal \cite{cabre07,2011MNRAS.414..329C}. For simplicity we consider an ideal full sky survey and use $\Delta \ell=1$ and $f_{sky}=1$. In this way we avoid correlations between different modes in the covariance matrix, which is diagonal with respect to $\ell$ (which is consistent with assuming the 3D covariance is also diagonal in $k$).

Therefore, for each $\ell$ we define a matrix with $N(N+1)/2$ elements, where $N$ is the number of observables discussed in Sec.~\ref{sec:2Dpower}, to account for the covariances and cross-covariances of auto and cross-correlations.
In order to include observational noise we add to the auto-correlations in Eq.~(\ref{eq:CovObs2D}) a shot noise term
\beq
C_{\ell}^{obs, ij}=C_{\ell}^{ij} +\delta_{ij}\frac{1}{\frac{N_{gal}(j)}{\Delta\Omega}}
\eeq
that depends on the number of galaxies per unit solid angle included in each radial bin.
We define the $\chi^2_{2D}$ assuming the observed power spectrum $C_{\ell}^{obs}$ correspond to our fiducial cosmological model discussed in Sec.~(\ref{sec:cosmologicalmodel}), while we call $C^{mod}_{\ell}$ the one corresponding to the cosmology being tested,
\beq
\chi^2_{2D}=\sum_{\ell} \left(C_{\ell}^{obs}-C_{\ell}^{mod}\right)^{\dagger}
{\rm Cov}_{\ell}^{-1}\left(C_{\ell}^{obs}-C_{\ell}^{mod}\right).
\label{eq:2Dchi2}
\eeq
Notice that each term in the sum is the product of $N_z(N_z+1)/2$-dimensional vectors $C^{ij}_{\ell}$ where $(ij)$ label all possible correlations of $N_z$ redshift bins, and
a $N_z(N_z+1)/2 \, \times N_z(N_z+1)/2$ matrix corresponding to their (inverse) covariance.

Recall that we use the exact calculation of $C_{\ell}$ using \camb, rather than the well-known Limber approximation \cite{1954ApJ...119..655L}.

\subsection{Nonlinear Scales}
\label{sec:kmax}

Both $\chi_{3D}$ and $\chi_{2D}$ depend sensibly on the maximum $k_{max}$ (or minimum scale) allowed in the
analysis. In this paper, we chose to fix $k_{max}$ for all the bins and relate it to angular scales through $\ell_{max} = k_{max} \, r(\zbar)$, where $\zbar$ is the mean redshift of the survey. In our fiducial cosmology we find $r(\zbar) = 1471 \Mpc$ in the redshift range $0.45<z<0.65$ and  $r(\zbar) = 2219 \Mpc$ when $0.4<z<1.4$. In addition, we do not consider a dependence of $l_{max}$ with redshift (i.e. same $\ell_{max}$ for all redshift bins and their cross-correlation).

For the largest scale we use $k_{min}=10^{-4}\kvecMpc$ in the 3D analysis and $\ell_{min}= 2 $ in the angular case. We have not found any significant dependence on $k_{min}$ or $\ell_{min}$.

\subsection{Cosmological model and growth history}
\label{sec:cosmologicalmodel}

We assume the underlying cosmological model to be a flat $\Lambda$CDM universe with cosmological parameters $w=-1$, $h=0.73$, $n_s=0.95$, $\Omega_{m}=0.24$, $\Omega_b=0.042$ and $\sigma_8=0.755$.
These parameters specify the cosmic history as well as the linear spectrum of fluctuations $P_0$.
In turn, the growth rate can be well approximated by,
\beq
f(z) \equiv \Omega_m(z)^\gamma
\label{eq:fz}
\eeq
and $\gamma = 0.545 $ for $\Lambda$CDM. Consistently with this we obtain the growth history as
\beq
D(z) \equiv \exp \left[ - \int_0^z \frac{f(z)}{1+z} dz    \right]
\eeq
(where $D$ is normalized to unity today). The parameter $\gamma$ is
usually employed as an effective way of characterizing modified
gravity models that share the same cosmic history as GR but different
growth history \cite{linder1}. In part of our analysis we
focus in $\Lambda$CDM models and assume the GR value $\gamma = 0.545$.
We deviate from this in Sec.~\ref{sec:biasfreecase} were we take
$\gamma$ as a free parameter independent of redshift.

\subsection{Likelihood analysis}

In order to find constraints on cosmological models
we integrate over the space of parameters defining the model, finding the value of the likelihood given by
\beq
-2\log \mathcal{L}\propto \chi^2,
\label{eq:likelihood}
\eeq
where we approximate the likelihood as Gaussian in the power spectra. Given the prior $\vartheta$ on the parameters one defines a probability for each sampled point $i$ in parameter space given by
\beq
{\mathcal{P}}(i) \propto \mathcal{L}(i) \times \vartheta(i).
\eeq
Finally, the mean and covariance matrix of the parameters is obtained from
\bea
\bar{p}_a & = &\sum_i{\mathcal{P}(i)p_a(i)} \label{eq:mean} \\
\indent \Sigma_{(p_a,p_b)}& = &\sum_i{\mathcal{P}(i)(p_a(i)-\bar{p}_a)(p_b(i)-\bar{p}_b)},
\label{eq:variance}
\eea
where $p_a(i)$ is the value of the parameter $a$ in the grid point $i$, $\bar{p}_a$ is the mean value and $\Sigma_{(p_a,p_b)}$ is the covariance between parameter $a$ and $b$. In Eqs.~(\ref{eq:mean},\ref{eq:variance}) $\mathcal{P}(i)$ is normalized to unity over the grid. In addition we assume flat priors.

By construction
%AL, see Eqs.~(\ref{eq:3Dchi2}) and (\ref{eq:2Dchi2}), the mean corresponds to 
the likelihood peaks at the fiducial value considered in the analysis. In all our studies we have chosen wide prior limits and therefore have found no dependance with these limits, and find the mean agrees with the fiducial value and the posteriors are quite Gaussian.
%AL as well as non-significant deviation from the gaussian assumption.
Then in the case of only one nuisance parameter $p$, solving $\chi^2(p)-1=0$ gives the same variance as likelihood sampling which allows us to speed up constraints considerably.

\begin{figure*}
\begin{center}
\includegraphics[width=0.7\textwidth]{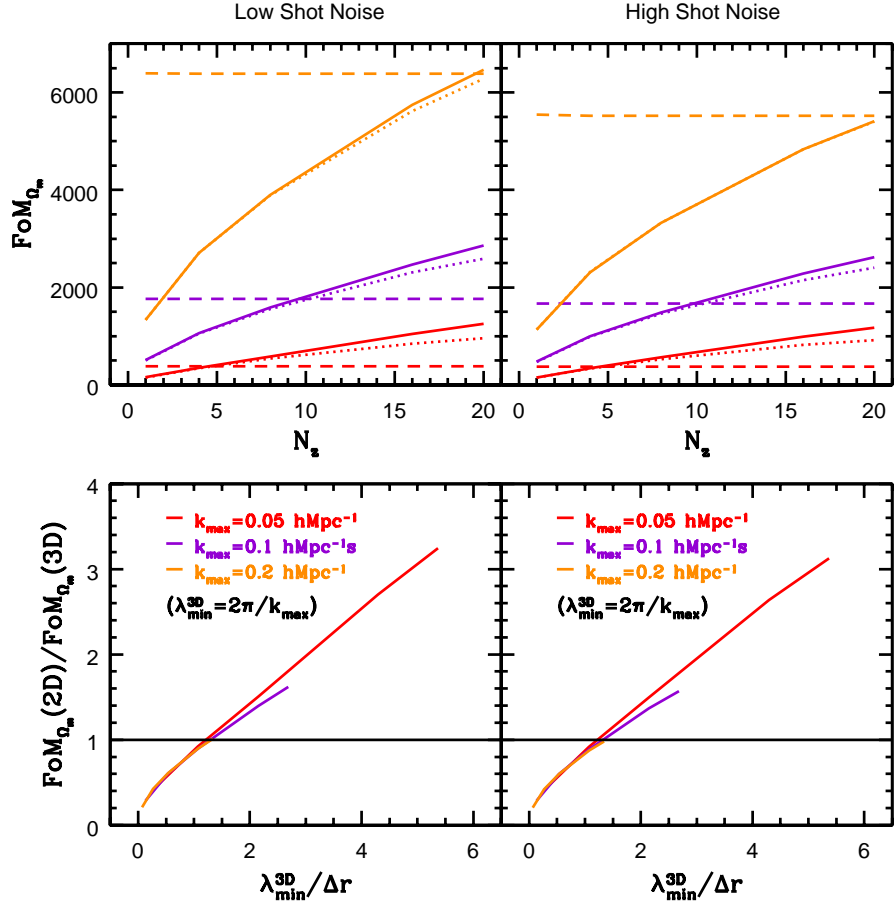}		
\caption{{\it Spectroscopic survey \& bias fixed}. Top panels show
  FoM$_{\Omega_m}$(2D) and FoM$_{\Omega_m}$(3D) as a function of the
  number of bins in which we divide the survey for the analysis (left
  panel for a low shot-noise survey and right to a high shot noise).
  Dashed line corresponds to the 3D analysis, dotted to the 2D
  tomography using only auto-correlations and solid to auto plus cross
  correlations. Different colors correspond to different minimum
  scales, as detailed in the bottom panel inset labels. Bottom panels show
  the ratio of FoM$_{\Omega_m}$(2D) (auto plus cross) and
  FoM$_{\Omega_m}$(3D) as a function of the bin width $\Delta r$ normalized by
  the minimum scale assumed in the 3D analysis. Remarkably the recovered
  constraints from full tomography match the 3D ones for $\Delta r \sim
  \lambda^{3D}_{min}$ for all $\lambda^{3D}_{min}$.
  We note that different lines in the bottom panels are truncated
  differently merely because we have done the three $k_{max}$ cases down to the same minimum
  $\Delta r$.}
\label{fig:FoMOmSPECRatios}
\end{center}
\end{figure*}

\subsection{Figures of Merit}
\label{sec:FoM}

We consider two different analyses in order to compare 3D clustering with 2D tomography including all the auto and cross-correlations between redshift bins.

On the one hand, a {\it bias fixed case}, in which we only vary $\Omega_{m}$ (which affects both the shape and the amplitude of the power spectrum, and can be constrained as if we had a good knowledge of the bias prior to the analysis).

On the other hand we consider a {\it bias free case}, in which only $b$
and $\gamma$ (hence $f$ through Eq.~(\ref{eq:fz})) are allowed to
vary. This changes the (anisotropic) amplitude of the power spectrum,
but not the underlying shape. This case is virtually the same as the
standard analysis of redshift space distortions
\cite{white2009,2011MNRAS.415.2193R}. For this case we had to adapt
\camb\, slightly, see the discussion in Appendix A.

To make the comparison quantitative we define a figure of merit (FoM) based
on the covariance matrix $\Sigma$,
\beq
FoM_{S}=\sqrt\frac{1}{{\rm det}[\Sigma]_S},
\label{eq:FoMdef}
\eeq
where $S$ is the subspace of parameters we are interested in. If this subspace correspond to only one parameter, then the FoM is the inverse of the square root of the variance of the corresponding parameter. Thus we have the following cases,

\begin{itemize}
\item{${\rm FoM}_{\Omega_m}$: Constraints on $\Omega_m$, with other parameters fixed at fiducial values.}
\parskip .2cm
\item{${\rm FoM}_b$ and ${\rm FoM}_\gamma$: bias and $\gamma$ constraints when marginalized over $\gamma$ and bias, respectively. Other parameters are fixed at their fiducial values.}
\parskip .2cm
\item{${\rm FoM}_{b\gamma}$: Joint constraint on bias and $\gamma$, with other parameters fixed at fiducial values}.
\end{itemize}

\section{Results}
\label{sec:results}

In this section we present the forecasts on $\Omega_m$  ({\it bias fixed}) and $b$ and $\gamma$ ({\it bias free}) from the measurement of either spatial or angular power spectra in the spectroscopic survey described in Sec.~\ref{subsec:zsurveys}. Next we perform the {\it bias fixed} analysis in the
narrow-band photometric survey with accurate photo-z discussed in Sec.~\ref{subsec:NBphotsurveys}
and the broad-band photometric survey defined in Sec.~\ref{subsec:BBphotsurveys}. Notice that despite photometric redshift errors  large scale redshift space distortions can be measured in photometric surveys for binned data \cite{2010MNRAS.407..520N,2011MNRAS.414..329C,2011arXiv1104.5236C}, albeit with possible large error bars. Nonetheless for photometric surveys
we concentrate the {\it bias fixed} case only.

All the analyses introduced above have been done for three different $k_{max}=\{0.05,0.1,0.2\} \kvecMpc$
(with corresponding $\ell_{max}$ as detailed in Sec.~\ref{sec:kmax}), and several bin configurations (see Table \ref{table:BinConfiguration} and \ref{table:BinConfigurationDES}). We then study for which redshift bin width the information obtained using angular power spectra (quantified by the FoM of Sec.~\ref{sec:FoM}) are similar to those derived from the 3D power spectra.

\subsection{Spectroscopic redshifts}

\subsubsection{Bias fixed case}

Top panels of Fig.~\ref{fig:FoMOmSPECRatios} show the FoM on $\Omega_m$ for different $k_{max}$ and $\ell_{max} = r(\bar{z}) k_{max}$
as a function of the number of redshift bins $N_z$ in which we divide the full survey volume (see Table~\ref{table:BinConfiguration}). Here dashed lines are results from fitting the 3D power spectrum according to  Eqs.~(\ref{eq:3Dchi2},\ref{eq:3Dtotalchi2}), while solid are from the 2D tomography including all the auto and cross correlations of bins, as in Eq.~(\ref{eq:2Dchi2}). The left (right) panel corresponds to the low (high) shot noise case as defined in Table \ref{table:Numgal}.

\begin{figure*}
\begin{center}
\includegraphics[width=0.7\textwidth]{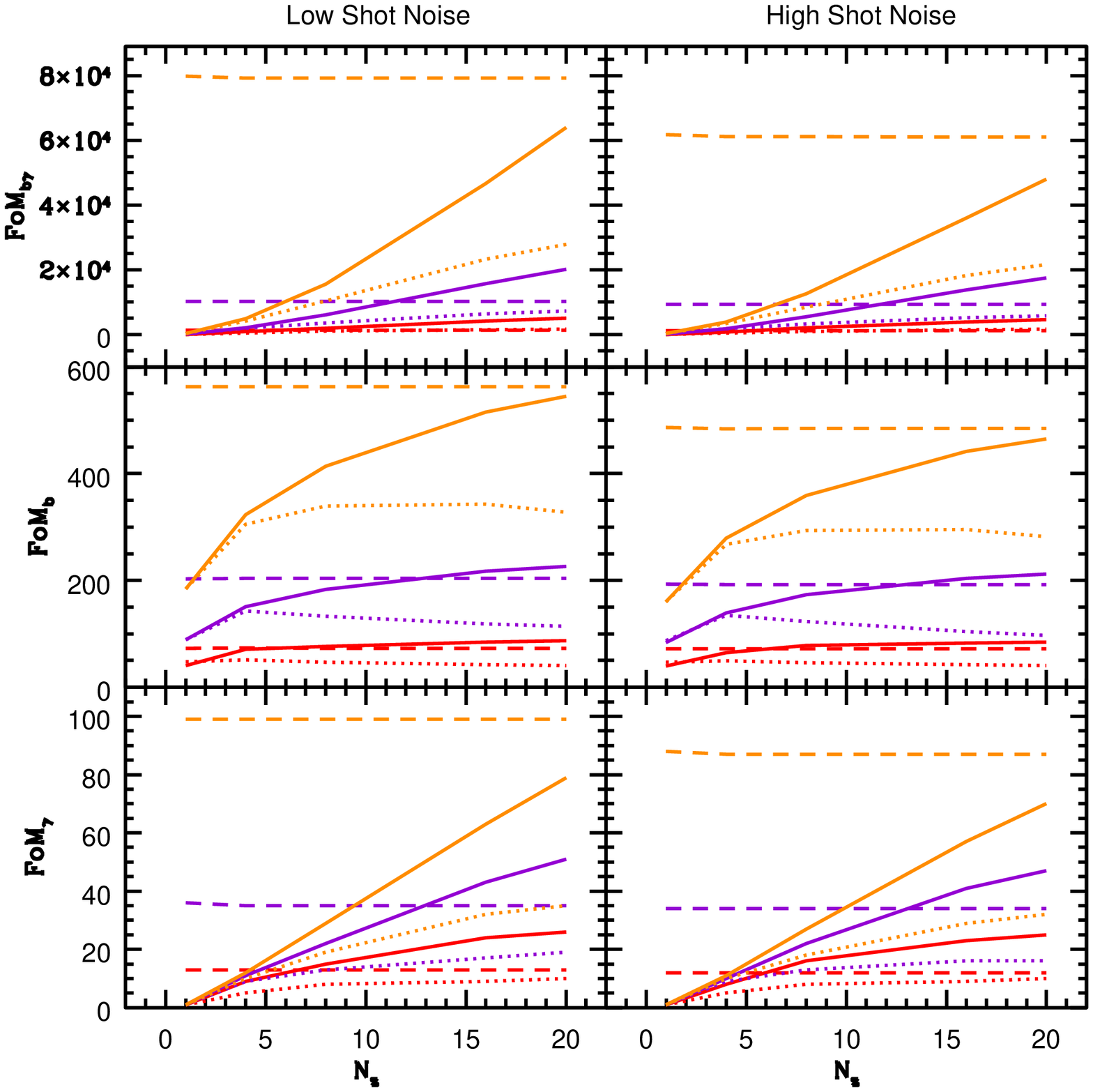}
\caption{{\it Spectroscopic survey $\&$ bias free}. Top panels show
  the combined $b-\gamma$ constraint resulting from 3D clustering
  (dashed lines) or 2D
  tomography considering as observables only auto
  correlations in redshift bins (dotted lines), or adding to this the
  cross-correlations (solid lines). The $x$-axis corresponds to
  the number of radial bins considered in the analysis. Different
  colors label different minimum scales assumed (same values and labels as in
  Fig.~\ref{fig:FoMOmSPECRatios}). Middle and bottom correspond to individual
  $b$ or $\gamma$ constraints after marginalization over $\gamma$ or
  $b$ respectively. As for the {\it bias fixed} we find that 3D
  information can be recovered but now the role of
  radial modes is much for important because RSD (our {\it bias free} case)
  relies on the relative clustering amplitude of radial and
  transverse mode.}
\label{fig:FoMbg}
\end{center}
\end{figure*}

As expected we find that the FoM increases for increasing $k_{max}$, $\ell_{max}$.
Including more modes to the $\chi^2$ adds more information to our analysis and therefore results in better constraints. We also see that FoM$_{\Omega_m}$ from the 3D analysis only show a marginal dependance on the bin configuration. This is because the $\chi^2$ per redshift bin is roughly proportional to the volume of the redshift shell, see Eq.~(\ref{eq:3Dchi2}). Thus, increasing the number of bins at the expense of decreasing their volume keeps the FoM$_{\Omega_m}$ unchanged.
We obtained the same result for all the cases studied in this paper, as long as $P_g$ does not change abruptly with redshift. Thus from now on we will only refer to the 3D results in the whole survey.

\begin{figure*}
\begin{center}
\includegraphics[width=0.8\textwidth]{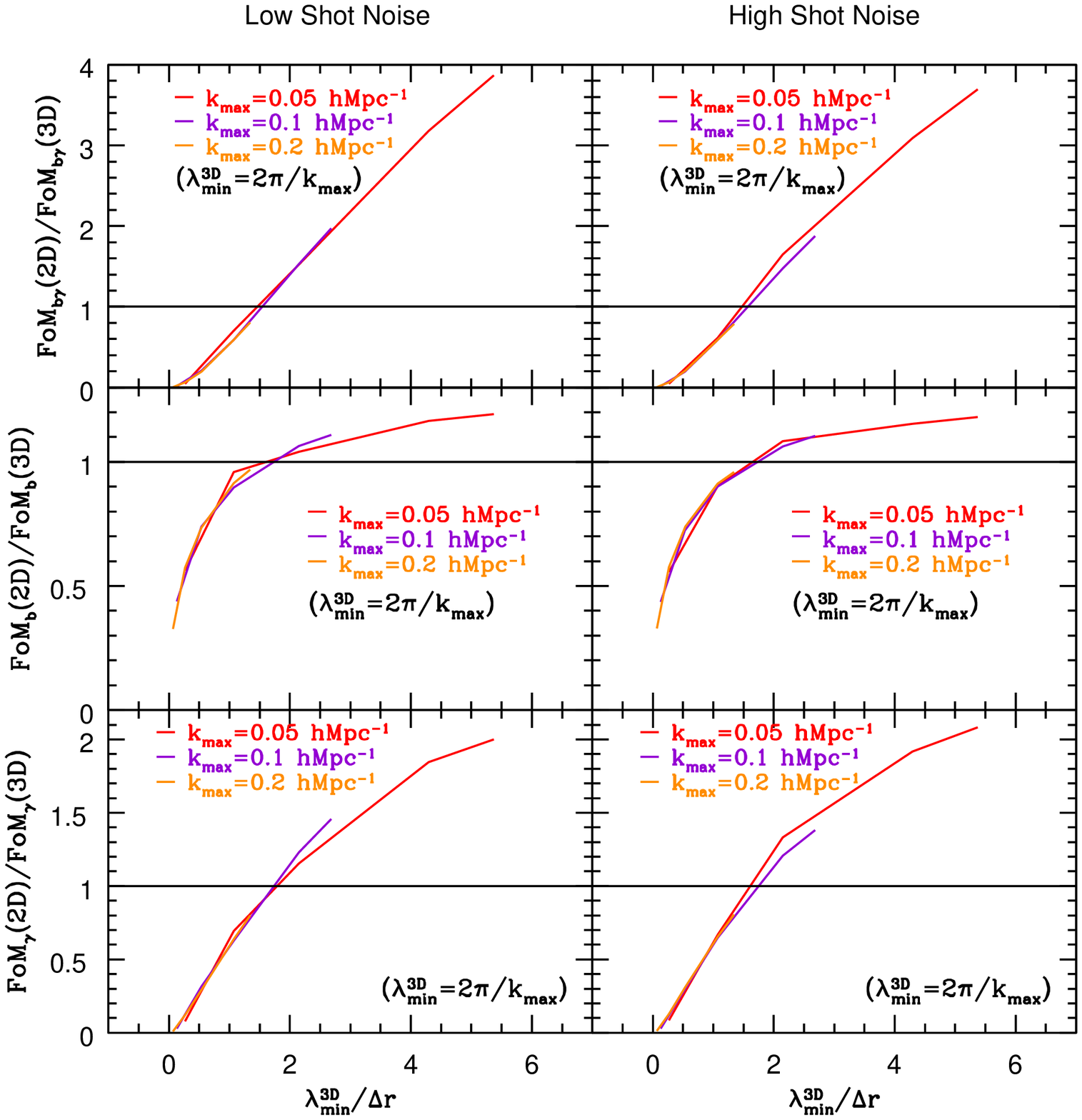}	
\caption{{\it{Spectroscopic survey \& bias free}}. Top panels show the ratio
  between combined FoM$_{b\gamma}$ (2D) (auto plus cross correlations) and FoM$_{b\gamma}$ (3D) with
  respect to $\lambda^{3D}_{min}=2\pi/k_{max}$, normalized by the mean
  width of the redshift bins $\Delta r$ in the analysis.
  Middle and bottom panels show the same but for ratios
  of FoM$_{b}$ and FoM$_{\gamma}$, respectively. We show results for
  Low Shot Noise and High Shot Noise in left and right panels,
  respectively. To reconstruct RSD information in practice, one need bins slightly smaller than $\lambda^{3D}_{min}$.}
\label{fig:FoMbgRatios}
\end{center}
\end{figure*}

This picture changes for the 2D tomography. Here the transverse
information is fixed once $\ell_{max}$ is set ($2\ell+1$ modes per
$\ell$ value up to $\ell_{max}$).
As we increase the number of
narrower bins $N_z$ (with fixed total redshift range)
we have several effects:

\begin{enumerate}
\item Decreasing the number of galaxies per bin increases the shot noise per bin
\newline

\item Increasing the number of bins so that they are thinner proportionally increases the signal auto power spectrum in each bin (there is less signal power suppression due to averaging along the radial direction).
\newline

\item When we split a wide redshift bin in two,
we double the number of angular auto power spectra (transverse modes). This results in a larger FoM because the signal to noise in each bin remains nearly constant (the shot noise and signal in each bin both increase proportionately). This gain is illustrated by the dotted line in
Fig.~\ref{fig:FoMOmSPECRatios},
which corresponds to the FoM produced by just using auto-correlations.
For even narrower redshift bins
the bins will become correlated and the gain will saturate, but this is not yet the case in our results as the redshift bins are still
large compared to the clustering correlation length. 
In the limit in which all modes of interest are
very small compared to shell thickness and they are statistically
equivalent, for a single power spectrum amplitude  parameter one expects FoM$=1/\sigma \propto \sqrt{N_z}$, as obtained in
Fig.~\ref{fig:FoMOmSPECRatios} for low $N_z$ \footnote{A similar effect can be seen on Figs. 8 and 9 of
    \cite{2011MNRAS.415.2193R} in the context of RSD constraints in a
    broad band photometric survey. In their Fig. 8 the
    constraint in $f\,\sigma_8$ saturates when they consider
    only one redshift bin. However the error on $f\sigma_8$ from the
    combined measurements on several bins does not saturate (Fig 9).}.
\newline

\item {When we increase the number of narrower bins,
we also include information of radial modes
by adding the cross-correlation between  different redshift bins
(illustrated by the solid line in Fig.~\ref{fig:FoMOmSPECRatios} that corresponds to the total FoM from auto plus cross-correlations).} Note how adding the cross-correlations
to the autocorrelations (solid lines in Fig.~\ref{fig:FoMOmSPECRatios}) only increases
the FoM moderately as compared to the autocorrelation result (dotted
line). This reflects the fact that there are fewer radial modes than
transverse ones, while much of the $\Omega_m$ constraint comes from
the shape of $P(k)$ that is isotropic.
\newline 

\item{As shown in  Fig.~\ref{fig:FoMOmSPECRatios}, the 2D FoM can exceed the 3D FoM.
This happens because the 3D analysis is limited by construction
to a maximum number of modes, given by $k_{max}$, while in 2D
we only limit the analysis to $l<l_{max}$ and we can formally exceed the maximum
number of narrow redshift bins,  as explained in point (ii) and (iii)
above. But in reality, these additional modes are not necesarilly independent and they 
could well be in the non-linear regime, so it is not clear to what extent we can use them to
increase the FoM. As we want to restrict our analysis to $k<k_{max}$
we should not use redshift bins that are smaller than $\lambda^{3D}_{min}$.}

\end{enumerate}

The bottom panels of Fig.~\ref{fig:FoMOmSPECRatios} show the ratio of the 2D and 3D FoM's against the bin width (instead of $N_z$), now normalized by the minimum scale used in the 3D analysis $\lambda^{3D}_{min}=\frac{2\pi}{k_{max}}$ (for three different $k_{max}$ as before).
We find FoM(2D) $\sim$ FoM(3D) when $\lambda^{3D}_{min} \sim \Delta r$
for all $\lambda^{3D}_{min}$. More precisely:
\begin{equation}
\Delta r = c \Delta z/H(z) \simeq 0.8 ~ \lambda^{3D}_{min}.
\label{eq:biasfixed}
\end{equation}

Basically this means that the 3D clustering information is recovered
once the binning is such that the radial bin width equals the minimum
scale probed in the 3D analysis. In this case one is able to constrain the parameters without loss of information compared to a three dimensional analysis,
though the actual range of scales around $k_{max}$ that
are used in the 2D analysis may be slightly different from the ones used in the
3D analysis.

Note that as mentioned in point (v) above, we can only really trust our results for
the 2D FoM up to the limit in which they are equal
or smaller than the 3D FoM, i.e., in the range in which the width of redshift bins is greater or similar than $\lambda^{3D}_{min}$. To use the smaller scales we first need to explore to
what extent we can model the non-linear 2D clustering to improve the FoM. We are currently investigating this issue (Asorey et al., in prep.)

\begin{figure*}
\begin{center}
\includegraphics[width=0.7\textwidth]{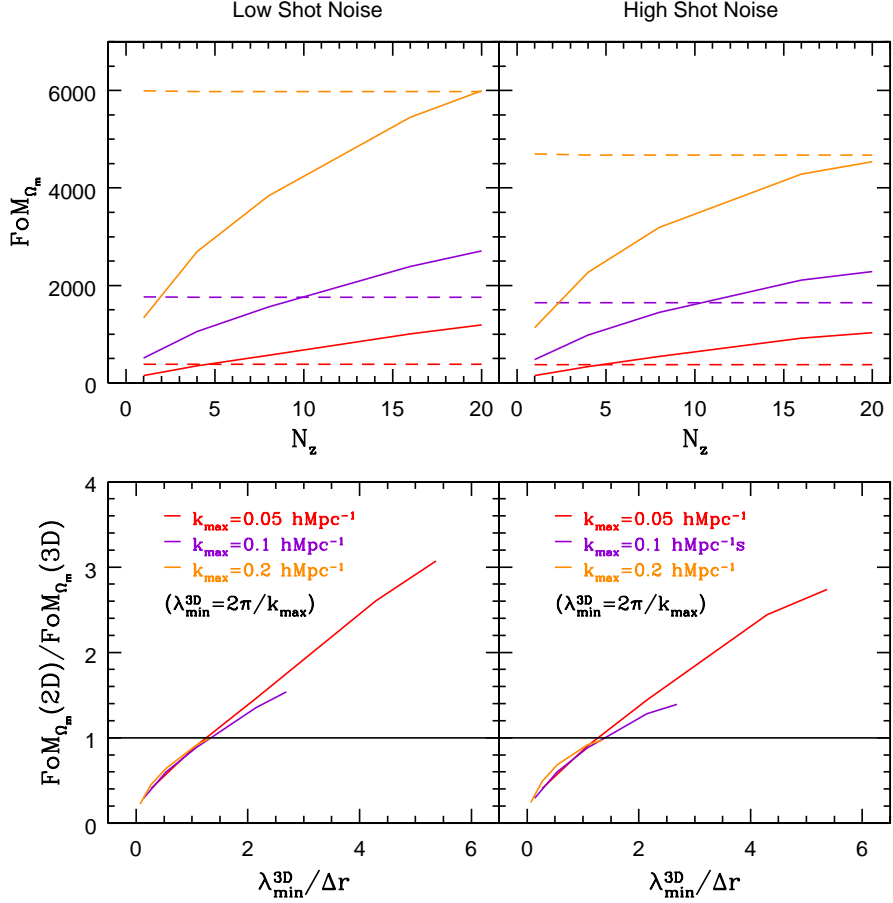}		
\caption{{\it Narrow band photometric survey (PAU-like) \& bias
    fixed}. Top panels show figures of merit FoM$_{\Omega_m}$ (2D)
  (auto plus cross correlations) and
  FoM$_{\Omega_m}$ (3D) with respect to the number of bins for
  $k_{max}=\{0.05,0.1,0.2\}$ hMpc$^{-1}$ (red, violet
  and orange colours). We plot 2D figures of merit with solid lines and
  3D figures of merit using dashed lines.  Bottom panels show the
  ratio between both figures of merit with respect to minimum scale
  used in 3D analysis, $\lambda^{3D}_{min}=2\pi/k_{max}$, divided by
  the mean width $\Delta r$ of the redshift bin. We conclude that we
  get similar constraints from 2D and 3D analysis when $\Delta r$ is
  close to $\lambda^{3D}_{min}$ and that in terms of bind width optimization an spectroscopic and photometric analysis are almost identical.}
\label{fig:FoMOmPAURatios}
\end{center}
\end{figure*}

Lastly, note that including shot noise does
degrade the FoM as shown in the right panel of
Fig.~\ref{fig:FoMOmSPECRatios}. However this does not change the
conclusions above.

\begin{figure*}
\begin{center}
\includegraphics[width=0.49\textwidth]{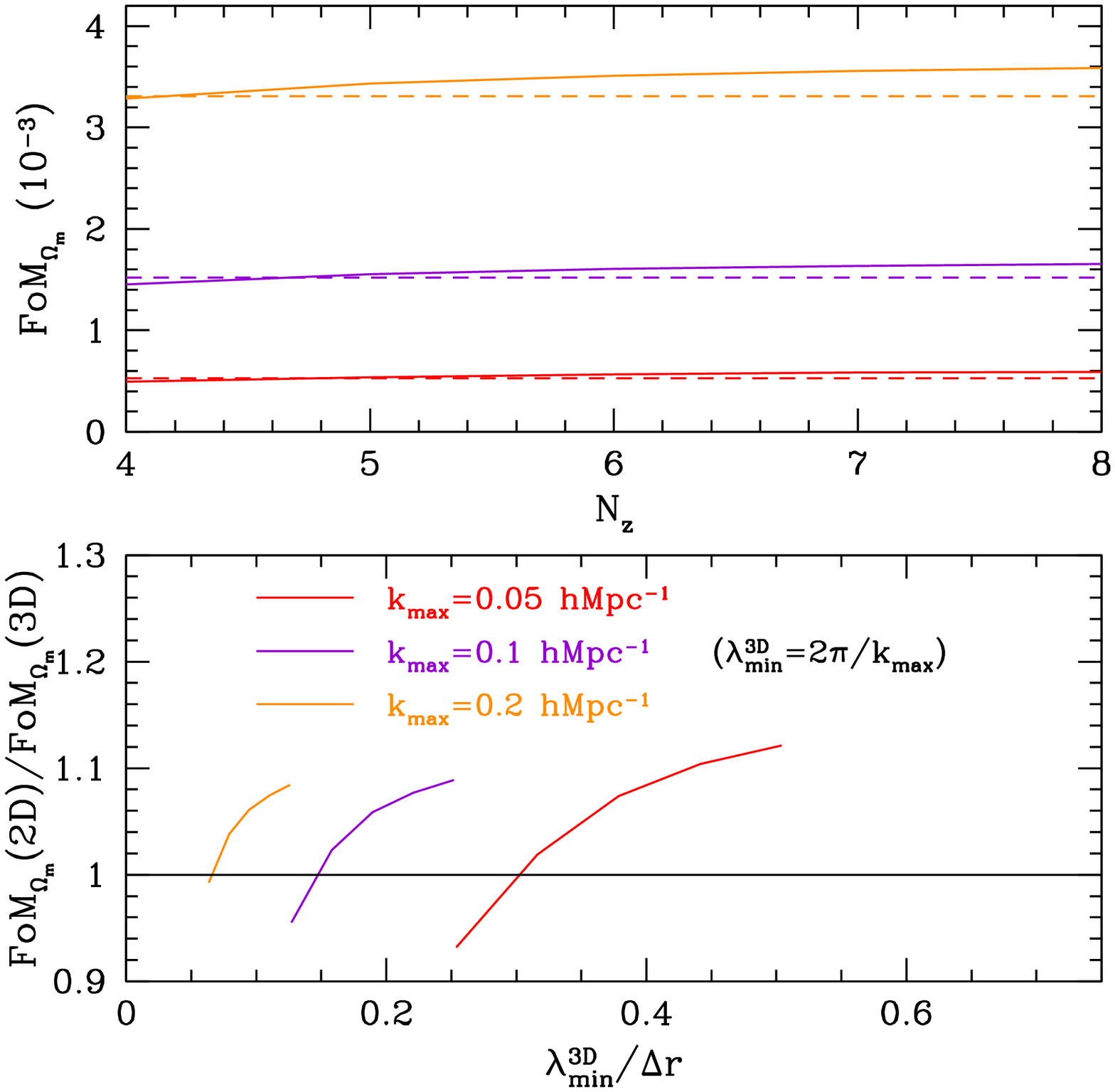}
\includegraphics[width=0.49\textwidth]{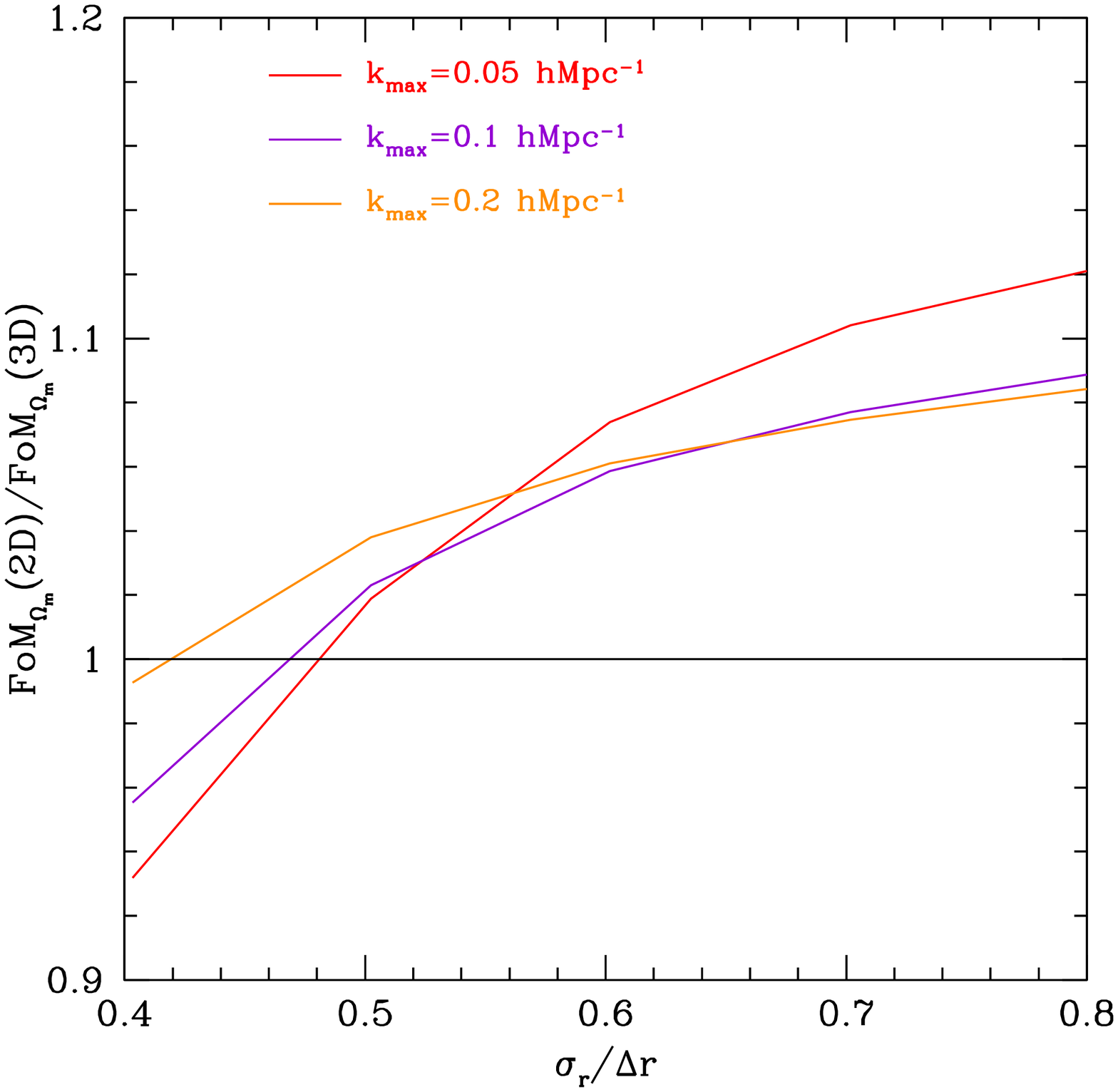}
\caption{{\it Broad band photometric survey (DES-like) \& bias
    fixed}. Top right panel shows the figures of merit
  FoM$_{\Omega_m}$ (2D) and FoM$_{\Omega_m}$ (3D) with respect to the
  number of bins $N_z$ used in the analysis,
  for $k_{max}=\{0.05,0.1,0.2\} \kvecMpc$
  (red, violet and orange colours, respectively). 2D FoM are plotted
  with solid lines and 3D with dashed lines, and we only consider low
  shot noise. Bottom left panel show the ratio of both figures of
  merit with respect to $\lambda^{3D}_{min}=2\pi/k_{max}$ divided by
  $\Delta r$. The equivalence of the recovered FoM now changes
  for different $k_{max}$. However when this ratio is plotted
  with respect to the comoving scale of photo-z, $\sigma_r$ (normalized by
  $\Delta r$) the different $\lambda_{min}$ lines cross each other for
  $\Delta r \sim 2 \sigma_r$. This implies that is the relative values
  of $\Delta r$ and $\sigma_r$ what sets the
  equivalence of 3D and 2D tomography.
   In particular, for a DES-like survey one recovers the 3D constraints
   from 2D analysis using 5 redshift bins.}
\label{fig:FoMOmDESRatios}
\end{center}
\end{figure*}

\subsubsection{Bias free case}
\label{sec:biasfreecase}

We now turn to the {\it bias free} case where we assume we know perfectly the shape of the power spectrum so that all the parameters are fixed at their fiducial values except the bias $b$ and the growth index $\gamma$.

In Fig.~\ref{fig:FoMbg} we plot the combined FoM obtained for bias $b$
and growth index $\gamma$, and the FoM of each of these 2 parameters
marginalized over the other, as a function of the number of redshift
bins considered in the analysis (for a fixed survey redshift range $0.45 < z
< 0.65$).
As in Fig.~\ref{fig:FoMbg}, dashed line corresponds to the 3D
analysis, dotted line to the 2D tomography using only auto-correlations\footnote{We note that we refer here to observables. The
  covariance of the auto-correlations does include cross-correlations
  of redshift bins, see Eq.~(\ref{eq:CovObs2D}).}
and solid line to the full 2D case where we add auto and cross angular
correlations.

We find a similar trend for the evolution of the different FoM of the
$\gamma$ and $b$ parameters (either combined or marginalized) than when varying $\Omega_m$.
Constraints given by spatial power spectrum are stable, while constraints
from projected power spectrum in the bins increases with the number of
bins in which we divide the survey. However there is a substantial
difference in regards to the contribution of radial modes. Now the
contribution of cross-correlations is very large (compare solid to
dotted lines in the left panel of Fig.~\ref{eq:CovObs2D}).
In fact, without cross-correlations we do not
recover all the 3D information.  This is because redshift space distortion information (i.e. our
{\it bias free case}) is based in the relative clustering amplitude of modes parallel and transverse to the line of sight. The
contribution from radial modes is much more evident for the $\gamma$
constraint (FoM$_\gamma$ and then FoM$_{b\gamma}$) because $\gamma$ is
basically what quantifies this relative clustering amplitude (in addition $f \equiv \Omega(z)^\gamma$ depends on redshift while we assume bias does not).

As we have done with FoM$_{\Omega_m}$ we show in
Fig.~\ref{fig:FoMbgRatios} the dependence of the ratios between 2D and
3D FoM with respect to $\lambda^{3D}_{min}/\Delta r$.
We find that both analyses produce the same constraints when the mean
redshift bin width is slightly smaller than
$\lambda^{3D}_{min}$  (and we use auto and cross 2D correlations).
Comparing these results with the {\it bias
  fixed} case, it seems that for the RSD probe we need to
extract more radial information. In this case:
\begin{equation}
\Delta r = c \Delta z/H(z) \simeq 0.6 ~ \lambda^{3D}_{min}
\label{eq:biasfree}
\end{equation}
as compared to 0.8 in  Eq.~(\ref{eq:biasfixed}).  This means that we have
to include more  radial bins when developing the fit to angular
correlations than when only fitting $\Omega_m$ if we want to match the
constraints  from 3D clustering.
This in practice corresponds to using slightly narrower redshift bins.
This may also result in more information being included from radial modes with $k>k_{max}$, though a detailed analysis of the implications of this is beyond the scope of the current paper.

\subsection{Photometric redshifts}

In this section we show how the results found in the previous section extend to the photometric surveys detailed in Sec.~\ref{subsec:NBphotsurveys} and \ref{subsec:BBphotsurveys}. For concreteness we will only consider the {\it bias fixed} study where all cosmological parameters are fixed at their fiducial values except for $\Omega_m$.

\subsubsection{Narrow-band photometric survey (PAU-like)}

In top panels of Fig.~\ref{fig:FoMOmPAURatios} we show the $\Omega_m$ constraints ({\it bias fixed} case) from 3D and 2D analysis (dashed and solid lines respectively) in  a narrow band photometric survey with $\sigma_z=0.004$.
In bottom panels we show how the ratio between 2D and 3D FoM depends on the ratio between the minimum scale of the 3D analysis and the mean comoving width of radial shells.

We find basically the same result as in the
spectroscopic survey. Constraints from a projected or unprojected analysis are equivalent when the mean width of the radial shells (set by our binning strategy) is equal to the minimum scale considered in 3D analysis $\lambda^{3D}_{min}$. The absolute value of each FoM is degraded with respect the FoM reached with an spectroscopic survey because photo-z errors dilute clustering in the radial direction. This broadens the selection functions in the 2D analysis and introduces a cut off already at quasilinear scales in the 3D $P(k)$. In both cases the consequence is that signal to noise reduces and thus errors of observables degrade. But if we compare Fig.~\ref{fig:FoMOmSPECRatios} and Fig.~\ref{fig:FoMOmPAURatios} we see that the spectroscopic survey and a photometric one with very accurate redshifts are almost indistinguishable in terms of bin width optimization.

%\subsection{Results for $\Omega_m$}
\subsubsection{Broad-band photometric survey (DES-like)}
We now consider a deep survey ($i_{AB}<24$) with redshifts estimated by
photometry with broadband filters ($\sigma_z=0.1$), and use the full catalogue with $0.4<z<1.4$. We obtain the FoM for $\Omega_m$ shown in the top left panel of Fig.~\ref{fig:FoMOmDESRatios}.

Now the large photo-z error removes most of the radial information, thus
all FoM$_{\Omega_m}$ are degraded with respect to spectroscopic and narrow-band photometric surveys. In addition, we find that FoM$_{\Omega_m}$ saturates with the number of redshift bins included in  the survey for every $k_{max}$. This effect is produced by the overlapping between true galaxy distributions at different bins induced by photo-z transitions.

We also find that the configuration in which spatial and projected analysis constrain $\Omega_m$ equally corresponds to the same number of bins for all the $k_{max}$ considered.
Therefore, as we can see in bottom left panel of
Fig.~\ref{fig:FoMOmDESRatios}, the scale given by
$\lambda^{3D}_{min}$ is not ruling the dependencies.  Instead it is
the scale of the photometric redshifts which is affecting both
clustering analyses. This is shown in the right panel of
Fig.~\ref{fig:FoMOmDESRatios} where we plot the ratio of figures of
merit (2D vs. 3D) against a new scaling : $\sigma_r/\Delta r$. We find
that for a DES-like case, with the assumption of $\sigma_z=0.1$, one
needs roughly $5$ bins for the 2D tomography to optimally recover the
3D clustering information. This corresponds to:
\begin{equation}
\Delta z \simeq 2 \sigma_z.
\end{equation}
With a lower $\sigma_z$ the number of bins will increase.

\section{Conclusions}\label{sec:conclusions}

In this paper we have studied the redshift bin width that
allows us to recover the full 3D clustering constraints from tomography of
angular clustering (i.e. the combination of all the auto and cross
correlations of redshift bins). We explore
three surveys with different properties: a spectroscopic and a narrow band photometric survey in a redshift range $0.45<z<0.65$, and a deeper broadband photometric survey that covers redshifts in the range $0.4<z<1.4$.  We have considered how well we can recover the shape of the power
spectrum by allowing  $\Omega_m$ to be free and fixing the amplitude of clustering, including
bias. We call this the {\it bias fixed} case.
We have also explored how to recover the information from redshift
space distortions (RSD), by measuring the {\it anisotropic} amplitude of the power spectrum allowing for
 both a free bias and a free growth index. This is the {\it bias free} case.
We restrict our study to quasi-linear scales and we only consider scales above some
minimum scale $\lambda^{3D}_{min}=2\pi/k_{max}$, where
 $k<k_{max}$ and $k_{max}$ is either 0.05, 0.1 or 0.2 $\kvecMpc$. In angular space
this corresponds to $l<l_{max} \simeq k_{max} r(z)$, where $r(z)$ is the radial distance
to the mean redshift bin.

The 3D analysis has almost no dependance on the number of redshift bins because radial modes are already included in each bin. In contrast the 2D tomographic analysis 
depends strongly on the number of bins (or equivalently on redshift bin widths), since broad bins average down transverse power on scales smaller than the bin width, and 
it is only by using multiple thin shells that radial modes are included.

For the {\it bias fixed} case in  the spectroscopic survey we have found that we recover all the
information with 2D tomography when the width of the redshift bins
that we use to do the tomography is similar to the minimum scale used
in the 3D observables, $\lambda^{3D}_{min}$. More precisely we find
that the optimal bin  width is (see Fig.~\ref{fig:FoMOmSPECRatios} and Eq.~(\ref{eq:biasfixed})):
$\Delta r = c \Delta z/H(z) \simeq 0.8 ~\lambda^{3D}_{min}$. In
addition most of the 2D constraints come from autocorrelations.

When studying RSD, i.e. in the {\it bias free} case,
we see that if we want to recover the 3D constraints we need radial shells which are slightly smaller, i.e. $\Delta r \simeq 0.6 \lambda^{3D}_{min}$ (see Fig.~\ref{fig:FoMbgRatios}),
which means that we would need more bins than in the case in which we
just want to measure the shape of $P(k)$.  In addition we find necessary
to include in the observables the cross correlation between redshift bins.
This is expected because in the RSD case we are comparing the
clustering in radial and transverse direction to the light of sight:
 information from radial modes should be more important than in the case in which we just study information in the isotropic shape of the power spectrum.
 Also note how we can not recover the 3D information from RSD
when we just use autocorrelations (see dotted line in Fig.~\ref{fig:FoMbg}).

We found that in the {\it bias fixed} case, the narrow-band
photometric survey is almost equivalent to an spectroscopic survey, and
we therefore reach the same conclusions  with respect to the optimal
bin width for the tomography of galaxy counts. In the case of a deeper
broadband photometric survey we find that the typical uncertainty in
photometric redshifts $\sigma_ z$
severely limits the accuracy of the radial information for both 3D and 2D cases. In this case the
information recovery  does not depend strongly on $\lambda^{3D}_{min}$, because this
is smaller than the scale corresponding to the photometric redshift
accuracy, i.e.
$c\,\sigma_z/H(z) > \lambda^{3D}_{min}$.
The optimal redshift bin width in this case is simply given by
$\Delta z \simeq 2 \sigma_z$.

For a redshift range $0.4<z<1.4$ and $\sigma_z=0.1$ (DES-like survey) we find that we will need only
5 redshift bins to  constrain $\Omega_m$ using tomography with the
similar precision than a full 3D analysis of the survey. In
comparison, for a PAU-like survey with $\sigma_z \simeq 0.004$ and $k_{max}=0.1$
we need about 44 redshift bins of width $\Delta z \simeq 0.023$ each.

We conclude from our analysis that it seems possible to recover the full
3D clustering information, including RSD information, from 2D tomography.
This has the disadvantage of needing a potentially large number
of redshift bins, and correspondingly large covariance matrices
between observables. But it
 has the great advantage of simplifying the combination with WL and of
just using observed quantities, i.e. angles and redshifts,
avoiding the use of a fiducial cosmology to convert angles and redshifts into
3D comoving coordinates. In practice, probably both types of analysis should be
used to seek for consistency.

\section*{Acknowledgments}
We thank Martin Eriksen and Pablo Fosalba for discussions and feedback on ideas in this paper.
J. A. would like to thank the great hospitality of the Department of Physics $\&$ Astronomy of the
University of Sussex, where part of this work was carried out. Funding for this project was partially provided by the
Spanish Ministerio de Ciencia e Innovacion (MICINN),
project AYA2009-13936,  Consolider-Ingenio CSD2007- 00060,
European Commissions Marie Curie Initial
Training Network CosmoComp (PITN-GA-2009-238356),
research project 2009-SGR-1398 from Generalitat de Catalunya
and  the Juan de la Cierva MEC program.
J. A. was supported by the JAE program grant from the Spanish National Science
Council (CSIC).
A. L. was supported by the Science and Technology Facilities Council [grant number ST/I000976/1].

\appendix

\section{Modifying CAMB$\_${\tt sources} to sample growth rate and bias}
In order to consider the {\it bias free} case we had to modify
\camb\,to accept as (independent) inputs bias and growth rate (parameterized through
$\gamma$ as in Eq.~(\ref{eq:fz})).
In addition this case does not involve changes in the shape of the
real space spectrum, thus one should be able to sample parameter space
without the need to compute the transfer functions at each point of
 parameter space.

To fulfil these needs we have factorized
the terms in our observables that depend on the cosmic history (for
our reference cosmology) from those that depend on the bias $b$ and growth index $\gamma$. The factorization in the case of auto and cross-correlation is given by:
\bea
\indent C_{\ell}^{ii}&=&b_i^2C_{\ell}^{ii\ (0)}+2b_if_iC_{\ell}^{ii\ (2)}+f_i^2C_{\ell}^{ii\ (4)}
\label{eq:RSDreconstructionAutoCl}\\
C_{\ell}^{ij}&=&b_ib_jC_{\ell}^{ij\ (0)}+b_if_jC_{\ell}^{ij\ (2)}\nonumber\\
&&+b_jf_iC_{\ell}^{ij\ (2')}+f_if_jC_{\ell}^{ij\ (4)}
\label{eq:RSDreconstructionCrossCl},
\eea
where $b_i$ is the bias of the bin $i$ and $f_i$ is the growth rate
given by Eq.~(\ref{eq:fz}), evaluated at the mean redshift of the bin
$i$. This factorization assumes $f(z)$ does not vary much within the
redshift range of the bin (neither $b$). We have tested this assumption using the
exact \camb\,evaluation or the reconstruction of Eqs.~(\ref{eq:RSDreconstructionAutoCl},\ref{eq:RSDreconstructionCrossCl}) and found an
excellent match for the bin widths considered in this paper.

Using the observed $C_{\ell}$ and solving a linear set of equations
using different values for $b_i$ we can store the value of
$%C_{\ell}^{ii\ (0)},C_{\ell}^{ij\ (0)},
C_{\ell}^{ii\
  (2)},C_{\ell}^{ij\ (2)},C_{\ell}^{ij\ (2')},C_{\ell}^{ii\ (4)}$ and
$C_{\ell}^{ij\ (4)}$. The values of $C_{\ell}^{ii\
    (0)}$and $C_{\ell}^{ij\ (0)}$ are obtaining by excluding RSD in $C_\ell$.
Then, we sample $b$ and $\gamma$
space using these factors and the reconstruction given by Eqs.~(\ref{eq:RSDreconstructionAutoCl}) and
(\ref{eq:RSDreconstructionCrossCl}) obtaining
$C_\ell^{mod}$ in parameter space.

In the reconstruction we assume the underlying value of
${\Omega_m}=0.24$ given by our reference cosmology while the growth
factor $D(z)$ is included in the integrals that are contained in the
cosmic history dependent factors $C^{ij\,(n)}_\ell$.

\bsp

\label{lastpage}

\end{document}